\documentstyle [epsf,multicol,aps,prl,epsfig] {revtex} 

\input epsf
 
\begin{document}
 \setlength{\unitlength}{1.0cm}
\twocolumn[\hsize\textwidth\columnwidth\hsize\csname
@twocolumnfalse\endcsname

\title{Comparison of averages of flows and maps}
\author{Z. Kaufmann$^1$\cite{email1} and 
H. Lustfeld$^2$\cite{email2}}

\address{$^1$ Department of Physics of Complex Systems, E\"otv\"os University,
P. O. Box 32, H-1518 Budapest, Hungary}
\address{$^2$ Institut f\"ur Festk\"orperforschung, 
  Forschungszentrum J\"ulich, D52425 J\"ulich, Germany}
\maketitle

\begin{abstract}%
It is shown that in transient chaos
there is no direct relation between averages in a continuos 
time dynamical system (flow)
and averages using the analogous discrete system defined by the
corresponding Poincar\'e map. In contrast to permanent chaos, 
results obtained from the Poincar\'e
map can even be qualitatively incorrect. The reason is that the return time
between intersections on the Poincar\'e surface becomes relevant. 
However, after introducing a true-time Poincar\'e map, 
quantities known from the usual Poincar\'e map, such as
conditionally invariant measure and natural measure, can be generalized
to this case. Escape rates and
averages, e.g. Liapunov exponents and drifts can be determined correctly
using these novel measures. Significant differences become evident
when we compare with results obtained from the
usual Poincar\'e map.
\end{abstract}

\vskip2pc] 
\narrowtext
Extensive investigations of chaotic systems in recent years have demonstrated
the great importance of transient chaos, due 
mainly to its connection with transport phenomena
\cite{GaNi90,GiFeDo99,TeVo00}
and chaotic advection
\cite{KaT},
possibly associated with chemical reactions
\cite{NeLoHeT}.
In most chaotic systems it is sufficient to know the intersection points of
the trajectories with a chosen surface $P$, the so-called Poincar\'e surface.
In case of $N$-dimensional phase space $P$ is $N-1$-dimensional.
Using a coordinate system on $P$,
and finding the connection between the successive intersections
$\mbox{\bf x}_n$ and $\mbox{\bf x}_{n+1}$
the Poincar\'e map ($PM$) can be constructed as
\begin{equation}
\mbox{\bf x}_{n+1}=\mbox{\bf f}(\mbox{\bf x}_n).
\end{equation}
The behavior of the system can then be studied by iteration of this map.
The advantages of the  use of $PM$ are (i) it is discrete, (ii) it has
smaller dimension. Its disadvantage is the absence
of the close connection between the number of 
intersections $n$ and the
time $t$, since the return time $\tau $ between two
intersections depends generically on where a trajectory intersects. 
One can keep this information by completing the $PM$ with the equation
\begin{equation}
t_{n+1} =t_n +\tau (\mbox{\bf x}_n)\;.
\end{equation}
We  call this extended map the {\em true-time} Poincar\'e map
($TPM$).

Usually, one reduces to the $PM$ by the following argument:
The total time after $n$ iterations is given by the sum
of the corresponding return times $\tau (\mbox{\bf x})$.
It is generally assumed that for large $n$ and for typical trajectories
the terms in the sum can be
replaced by their average over the invariant density $\rho_P$ of the map.
The sum then becomes a product
\cite{EckmRuel85}
\begin{equation}
t = n \langle \tau \rangle ,\;\; \langle \tau  \rangle 
=\int_P dx\,\rho_P(\mbox{\bf x}) \tau (\mbox{\bf x}),
\label{nConnPM}
\end{equation}
Based on this connection, averages of the map (using $n$ for time) and
the flow (using the real time $t$) 
would be simply related by a time scale.
This is explicitly shown for general averages in case of
permanent (non-transient) chaos \cite{Gasp96}.

We demonstrate in this paper that, in contrast to {\em permanent} chaos,  
the situation is quite different for {\em transient} chaos.
Not only $\langle\tau\rangle$ in Eq.\ (\protect\ref{nConnPM}) 
should be changed,
but averages of the map and the flow (or of the $TPM$ representing it)
are not any more related by a time scale.
The situation is somewhat reminiscent of the case when, instead of simple
averages, the decay rates of correlations are considered.
Even in permanent chaos, these
show a discrepancy in non-ideal situations \cite{Buni85}.
To proceed in a correct manner we must
start with the
$TPM$, which contains all the information needed for the longtime 
behavior of the system and from which we can derive all necessary
formulas. Finally we compare these with the corresponding
ones of the $PM$ by setting $\tau(\mbox{\bf x}) =\langle \tau\rangle$.
The use of the $PM$ is sufficient if the results
do not change.

It is convenient to initiate the trajectories by
inserting particles on $P$ with an input current density
$\rho_{in}(\mbox{\bf x},t)$.
Let $\rho_{in}(\mbox{\bf x},t) \equiv 0$ for $t <0$.
Since a trajectory leaving $P$ has either been
initiated  there or has intersected $P$ previously, we obtain for 
the normal component $\rho_P(\mbox{\bf x},t)$ of the current density on $P$
\begin{equation}
\rho_P=({\cal L}\rho_P) + \rho_{in}\;. \label{eqRho}
\end{equation}
Here ${\cal L}$ is the Frobenius-Perron operator of the $TPM$
which is defined by

\begin{equation}
({\cal L}g)(\mbox{\bf x},t) =\int_P d\mbox{\bf x}'
\delta(\mbox{\bf x} -\mbox{\bf f}(\mbox{\bf x}'))
g(\mbox{\bf x}',t-\tau(\mbox{\bf x}'))\;. \label{timeFrob}
\end{equation}

Quite often the motion in one direction --- the unstable one ---
depends at most weakly on the others.
Choosing a
coordinate system
in which $x$ is taken along this direction evolution of $x$
can be well approximated by a one-dimensional map $x_{n+1}=f(x_n)$.
(The price paid is the non-uniqueness of $f^{-1}$).
This happens, for example, in strongly dissipative systems and
in those analogous to Baker type maps. 
In such a situation Eq.\ (\protect\ref{eqRho}) remains valid
if $\mbox{\bf x}$ is replaced by $x$ and $\mbox{\bf f}$ by $f$,
and projecting the densities onto the unstable direction.
For simplicity we restrict our attention to this one-dimensional case.

First we compute the quasistationary distribution. 
We assume that the system will become 
quasistationary after some time, i.e. that
the distribution  
decays exponentially but all relative weights remain
constant.
Normalization of the distribution leads  to the time-independent
conditionally invariant density\cite{PiYo79,PiYo81,Tel90} $\rho_P(x)$.
We make the ansatz
$\rho_P (x,t) =\rho_P(x) e^{-\kappa t}$,
where $\kappa $ is the escape rate,
and obtain from Eq.\ (\protect\ref{eqRho}) the selfconsistent 
equation for the conditionally invariant density 
\begin{equation}
\rho_P(x) =\int_I\!dx'\delta (x -f(x'))e^{\kappa \tau (x')}\rho_P (x')\;.
\label{cid}
\end{equation}
Here $I$ is the range of the values of $x$. 

For example, 
let $f$ be the tent map with a possible
opening:
$f(x)=x/a_0$ if $x<a_0$,
$f(x)=(1-x)/a_1$ if $x>1-a_1$,
where $a_0+a_1\le 1$.
Furthermore let $\tau(x)$ be piecewise constant:
$\tau(x)=\tau_0$ if $x<a_0$,
$\tau(x)=\tau_1$ if $x>1-a_1$.
Escape occurs for $a_0+a_1<1$, when the trajectory leaves the Poincar\'e
surface in the interval $x\in(a_0,1-a_1)$.
The smooth, non-negative solution of (\ref{cid}) is now
$\rho_P(x)=1$. By chance $\rho_P(x)$ does not depend on $\tau(x)$,
but the equation
for $\kappa $ does:
\begin{equation}
a_0 e^{\kappa\tau_0} +a_1 e^{\kappa\tau_1} =1\;,
\end{equation}
and in nonlinear maps $\rho_P(x)$ also depends on it.

For a general treatment we write the formal solution of
Eq.\ (\protect\ref{eqRho}) as
\begin{equation}
\rho_P(x,t)=[({\bf 1} -{\cal L})^{-1}\rho_{in}](x,t)\;,
\label{rhosol}
\end{equation}
We continue with Laplace transformations in time,
since
the generalized operator can be written as
$({\cal L}g)(t) =\int_0^t dt'\,L(t-t')g(t')$,
with
$L(\Delta t)g(x,t')=\int_I dx'\,
\delta(f(x')-x)\delta(\tau(x')-\Delta t)g(x',t')$. Its 
Laplace transform (denoted by $\tilde{ }$ ) is 
$\widetilde{{\cal L} g}=\tilde L\tilde g$ and 
Eq.\ (\ref{eqRho}) yields
\begin{equation}
\tilde\rho_P =(1-\tilde L)^{-1}\tilde \rho_{in}\;.\label{rhoinl}
\end{equation}
Considering $s$ as a parameter we can use the eigenfunctions satisfying
\begin{equation}
\tilde{L}(s)\varphi_m(s)=\lambda_m(s)\varphi_m(s)
\label{eigenValEq}
\end{equation}
to expand $\tilde{\rho}_{in}$ as
$\tilde{\rho }_{in}(s) =\sum_0^\infty a_m(s)\varphi_m(s)$.
Inverse Laplace transformation gives
$$
\rho_P(x,t) =\frac{1}{2\pi i}\int_{c-i\infty}^{c+i\infty}ds\, 
\sum_0^\infty 
\frac{e^{st}}{1-\lambda_m(s)} a_m(s)\varphi_m(x,s)\;.
$$
Each value of $s$ for which $\lambda_m(s)=1$ with some $m$ gives a pole in
the integrand and a term $e^{s t}$ in $\rho_P(t)$.
Therefore, the leading asymptotic time dependence is $e^{-\kappa t}$,
and hence the escape rate $\kappa$ is determined by the position of the
leading pole,
i.e.
\begin{equation}
\kappa =-s_0,\; s_0 =\max_{m}\{s\mbox{ with } \lambda_m(s)=1
\mbox{ and $s$ real}
\},\label{pole0}
\end{equation}
where we assume for simplicity that $s$ is maximal for $m =0$.
($s$ must be real, otherwise $\rho_P(t)$ could not remain positive for
all $t$.)
Eq.\ (\protect\ref{eigenValEq}) together with Eq.\ (\ref{pole0})
corresponds to (\ref{cid}),
however, here we have obtained the result
and the decay
\begin{equation}
\rho_P(x,t) \approx\frac{1}
{-\lambda_0'(-\kappa )}e^{-\kappa t} a_0\varphi_0(x,-\kappa )
\label{condInvMeas}
\end{equation}
for large times without prescription of $e^{-\kappa t}$.

Typical long time averages are determined by trajectories
in the following manner. The quantities
in question consist 
for each trajectory of a sum of
terms $A(x)$ that are added whenever the trajectory
intersects $P$, i.e. we need 
averages of expressions of the form $\sum_l A(x_l)$ 
with $x_l =f^l(x_0)$. We consider some examples: 
If we are interested in the number of intersections $n$, 
we set $A(x) \equiv 1$, for the total time 
we set $A(x) =\tau (x)$. If we are interested in the leading 
Liapunov exponent
describing the exponential deviation of infinitesimally
close trajectories
we need the logarithm of the full derivative of $f$,
i.e. we must set 
$A(x) =\ln \mid f'(x)\mid $.

We now outline the case of the ordinary $PM$.
The average of $\sum_l A(x_l)$ 
is obtained by taking into account the contribution of 
all trajectories present after $n$ iterations and dividing
this by the weight of these still present trajectories. Thus we arrive at
$\int_I dx\int_I dx_0\,\delta(x_n -x)
\sum_{l =0}^{n-1}A(x_l)\rho_{in}(x_0)/(\int_Idx\,\rho _P(x,n))$. 
The denominator is an integral of the density
$\rho_P(x,n) =\int_I dx_0\,\delta(x_n -x))\rho_{in}(x_0)$. Correspondingly
we can write the numerator as an integral of a 'weight
density'  $\sigma_A(x,n) =\int_I dx_0\,\delta(x_n -x)
\sum_{l =0}^{n-1}A(x_l)\rho_{in}(x_0)$.

The $TPM$ requires two modifications:
(i) instead of $n$ we take a time interval of length $t$,
(ii) an additional sum over the number of
intersections $n$ is necessary,
since $n$ is generically 
different for different trajectories.
This yields for the weight density
\begin{eqnarray}
\sigma_A(x,t)&=&\sum_{n =0}^\infty 
\int_I dx_0\,\int
dt_0\,\delta(x_n -x)\\
&\times&\delta(t-t_0 -\sum_{j =0}^{n-1}\tau(x_j))
\sum_{l =0}^{n-1}A(x_l)\rho_{in}(x_0,t_0)\, ,\nonumber
\end{eqnarray}
where $\delta(t-t_0 -\sum_{j=0}^{n-1}\tau(x_j))$ demands that the 
trajectories intersect $n$ times during time $t -t_0$.
The average value of $\sum_l A(x_l)$ is given asymptotically as
\begin{equation}
t \langle A \rangle_\mu  =t\lim_{t'\rightarrow\infty}
\frac{1}{t'}\frac{\int_I dx \sigma_A(x,t')}{\int_I dx \rho_P(x,t')}
\end{equation}
The index $\mu $ signalizes that asymptotically this average 
{\em does not} depend on $\rho _{in}$ but on the natural 
measure $\mu $ of the $TPM$. This will now be explained.

To compute $\sigma_A $ we note first that
\begin{equation}
\sigma_A ={\cal L}(\sigma_A +A\rho_P )\;.\label{shiftDens}
\end{equation}
This is analogous to Eq.\ (\ref{rhosol}).
Defining ${\cal T} =(1-{\cal L})^{-1}$
we obtain
\begin{equation}
\sigma_A ={\cal T}{\cal L} A {\cal T}\rho_{in}\;.\label{sigsol}
\end{equation}
We introduce $T$ by $({\cal T}g)(t) =\int_0^t dt'\,T(t-t')g(t')$,
similarly to the connection of ${\cal L}$ and $L$.
We can write the Laplace transform of (\ref{sigsol}) in terms of
the adjoints $\tilde{L}^+$ of $\tilde{L}$ 
and $\tilde{T}^+$ of $\tilde{T}$.
For the backward transformation of this expression we need the
eigenfunctions of $\tilde{L}^+$.
The solutions $\psi_m(s)$ of
$\tilde{L}^+(s)\psi_m(s)=\lambda_m^*(s)\psi_m(s)$
are functionals\cite{Rugh92,KaGr85,Gasp92}
(and can be approximated with strongly oscillating
functions), due to the fractal nature of the invariant set.
We insert an expansion $1=\sum_0^\infty b_m \psi_m$
and observe that, for large $t$, the most important terms occur 
when poles induced by
$\tilde{T}^{+*}(s)$ and $\tilde{T}(s)$ coincide.
Thus we obtain for large $t$
\begin{eqnarray}
\int_I\!\!dx\,\sigma_A(x,t)
\!=\!\frac{a_0 b_0 t e^{-\kappa t} }{\lambda_0'^2(-\kappa )}
\!\!\int_I\!\!dx\,\psi_0(x,-\kappa)
A(x) \varphi_0(x,-\kappa).\!\!\label{sigmaSol}
\end{eqnarray}
(Note that both $\psi_0(-\kappa )$ and
$\varphi_0(-\kappa )$ are real.) 
The prefactor of Eq.\ (\protect\ref{sigmaSol}) can be expressed 
by setting $A(x)=\tau(x)$ because of the relation, valid for large times,
$\langle \tau \rangle_\mu  =1$. 
Asymptotically we obtain
\begin{equation}
\langle A\rangle_{\mu} =\frac{1}{\langle \tau \rangle}
\int_I\! dx\,\psi_0(x,-\kappa ) \varphi_0(x,-\kappa )
A(x)\, ,
\label{muAver}
\end{equation}
where $\langle \tau \rangle \equiv
\int_I dx\,\psi_0(x,-\kappa ) \varphi_0(x,-\kappa )
\tau(x)$ with the normalization 
$\int_I dx\,\psi_0(x,-\kappa ) \varphi_0(x,-\kappa )
=1$.
Since these relations are valid for every observable $A$, we identify 
the natural
measure\cite{KaGr85} for infinitesimal intervals as

 \begin{figure}[htb]
 \begin{picture}(8.0,3.8)
 \put (-0.5,0.2){\epsfig{file =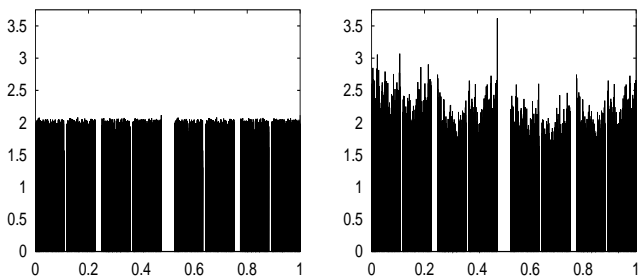,scale =0.15,
       height =4.0cm,width =9.0cm,angle =0}}
 \end{picture}
 \caption{Fractal distributions of the natural measures
 when modeling $\mbox{\bf f}(\mbox{\bf x})$ by the open tent map
 ($a_0=a_1 =0.475$)
 left: normal Poincar\'e map (scale $n$), $\tau_0 =\tau_1 =1$. right: 
 true-time Poincar\'e map (scale $t$),
 $\tau_0 =1$, $\tau_1 =0.1$.}
 \label{fig1}
 \end{figure}

\begin{equation}
\mu([x,x+dx))=\frac{1}{\langle \tau \rangle}
\psi_0(x,-\kappa ) \varphi_0(x,-\kappa )\,dx\;.
\end{equation}
A comparison between the natural measure of the $PM$ and the $TPM$ 
for the tent map (Fig.1) shows obvious differences. It is clear
that the $repeller$ of the $PM$ and $TPM$ differ significantly,
(although their dimension $D_0$ is the same.)

The Liapunov exponent 
can be written as
\begin{equation}
\lambda_{\rm Liap}=
\int_I d\mu_P\, \ln \mid f'(x)\mid \label{liapAv}\;.
\end{equation}

For $A(x) \equiv 1$ we find $\bar n(t)=t\langle 1 \rangle_\mu 
=t/\langle\tau\rangle$,
thereby
\begin{equation}
t =\langle \tau \rangle \bar n \label{nConnT}\;.
\end{equation}
If we compare Eq.\ (\protect\ref{muAver}), Eq.\ (\protect\ref{nConnT}) and 
Eq.\ (\protect\ref{liapAv}) with Eq.\ (\protect\ref{nConnPM}) we 
see a complete analogy,
except that $\tau $ has to be averaged
over a quantity of the $TPM$.
This means that the flow of a repelling system is well represented by its 
$TPM$ and {\em not} by its $PM$.

The leading Liapunov exponent for the repelling tent map is
\begin{equation}
\lambda_{\rm Liap} =\frac{1}{\langle\tau\rangle}
\frac{a_0\exp{\kappa\tau_0}\ln (a_0^{-1}) 
+a_1\exp{\kappa\tau_1}\ln(a_1^{-1})}{
a_0\exp{\kappa\tau_0}+a_1\exp{\kappa\tau_1}}.
\end{equation}
In this example we see the irrelevance of $\tau $ for 
$\kappa  =0$ (no transient chaos), so that
$\langle \tau \rangle$ only sets the time scale. On the other hand, 
for $\kappa  >0$
$\lambda_{\rm Liap}$ 
is {\em not} invariant to changes of $\tau_0$
relative to $\tau_1$, proving again that $\tau $ is a
relevant quantity in transient chaos.

The comparison of the behavior of
the flow and the map shows that the occurrence of criticality can change when
turning from the map to the real system.
A state of a system is called critical if the natural measure is
concentrated on a subset of the repeller, while the invariant measure is
distributed on the whole repeller.
In such situations
there are two conditionally invariant measures
with different escape rates
\cite{NeSz95,LuSz96,KLNSz}.
Here we use the piecewise parabolic 1D map\cite{NeSz95}
that is defined on the interval $[0,1]$
by its inverse branches
$f_l^{-1}(x) ={(x+d\cdot x(1-x))}/{2R}$ (lower branch),
$f_u^{-1}(x) =1-f_l^{-1}(x)$ (upper branch)
and choose $\tau(x)=1+\tau_x\cdot(x-1/2)$.

 \begin{figure}[htb]
 \begin{picture}(8.0,3.8)
 \put (0.9,0.2){\epsfig{file =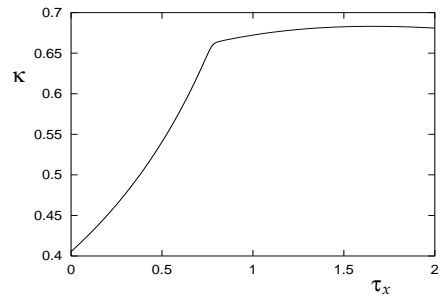,scale =0.15,
       height =4.0cm,width =6.0cm,angle =0}}
 \end{picture}
\caption{
The leading escape rate of the piecewise parabolic map as a function of 
the derivative of
$\tau(x)$,$\tau_x$.}
 \label{fig2}
\end{figure}
When increasing $\tau_x$ at a certain value the escape rates
change order of the modulus.
This is evidently a breakpoint in Fig.\ 2,
which shows the leading (smaller) one of the escape rates.
Above that point criticality disappears.
This will be explored in more detail elsewhere.

If P has a periodic structure it can be reduced to a unit cell with
periodic boundary conditions.
In this case both $PM$ and $TPM$ can be separated 
into a reduced map (which maps
the unit cell into itself) plus a shift 
$\Delta(x)$ describing the transit between the cells
\cite{GrFu82,GeNi84}.
Such systems can be characterized by the drift speed and diffusion
coefficient.
If particles can be lost from the point of view of diffusion by
absorption, chemical reaction or escape in directions transverse
to the extension of the system, we refer to transient diffusion
\cite{KLNSz,CMar,ClGa}.
We set $A =\Delta $ and we obtain a shift density
$\sigma_\Delta$
in analogy to the procedure above: 
$\sigma_\Delta  ={\cal T}{\cal L}\Delta {\cal T}\rho_{in}$.
We then determine the drift speed as the normalized shift per time
\begin{equation}
v =\lim_{t\rightarrow\infty}\frac{ \mbox{ average of }\{S(t)\}}{t}
=\frac{1}{\langle \tau  \rangle}
\langle \Delta\rangle_\mu\;.
\label{deltaAv}
\end{equation}

As an example of the essential role of averages over the natural 
measure we consider a diffusive
system on a one-dimensional lattice and assume that the reduced map
is the tent map. We assume
furthermore a microscopic process determining whether and by how much
a particle jumps to the left or to the right on the lattice.
We consider a lattice with period $1$
and a change of the coordinate by $\pm 1$ depending on the location in
the subinterval $[0,1]$, namely
$\Delta(x)=-1$ if $x\in[0,a_0]$,
$\Delta(x)=1$ if $x\in[a_1,1]$.
To calculate the average speed or the diffusion 
coefficient an average over
long trajectories is required. The average speed is
\begin{equation}
v
=\frac{1}{\langle \tau \rangle}
\frac{a_1\exp{\kappa\tau_1}-a_0\exp{\kappa\tau_0}}{
a_1\exp{\kappa\tau_1}+a_0\exp{\kappa\tau_0}}\;.
\end{equation}
Again if $\kappa  =0$ (non transient chaos) the return time $\tau $ sets the
time scale only, and $PM$ and $TPM$ give the same result.
But for $\kappa  >0$ (transient chaos)
even the sign of the speed 
can change when computing it with the usual $PM$, i.e. when setting
$\tau(x) \equiv\langle \tau \rangle \equiv const$.
Results for the diffusion coefficient will be published elsewhere.

In conclusion we have shown that the return time $\tau $ i.e. the time
between two successive intersections on the Poincar\'e surface $P$,
is a {\em relevant quantity in transient chaos}. The 
usual Poincar\'e map does not reflect the long time
averages of the flow 
satisfactorily and can even be completely
misleading. The solution is 
to use a true-time Poincar\'e map $TPM$
and its generalized Frobenius-Perron operator (\ref{timeFrob}),
where we can also define conditionally
invariant measure and natural measure. Escape rate,
Liapunov exponents, drift speed,
etc.\ depend significantly on $\tau(x)$
and are described correctly only by using the $TPM$. 
Therefore the necessary generalization of
the normal Poincar\'e map is the true-time Poincar\'e map
if the system in question is a repeller.

We thank J. Bene, R.O. Jones and G. Eilenberger 
for helpful discussions and comments on the manuscript.
This  work has been supported in part by the 
Hungarian National Scientific Research Foundation under Grant No.
OTKA T017493. Z.K. thanks G. Eilenberger
for the hospitality of the IFF, Forschungzentrum Julich, where 
part of this work was done.


\begin{thebibliography}{29}
\vspace*{-7mm}
\bibitem[*]{email1} Electronic address: kaufmann@complex.elte.hu
\bibitem[+]{email2} Electronic address: h.lustfeld@fz-juelich.de

\bibitem{GaNi90} P. Gaspard, G. Nicolis,
  Phys.\ Rev.\ Lett.\ {\bf 65}, 1693 (1990).
\bibitem{GiFeDo99} T. Gilbert, C. D. Ferguson, and J. R. Dorfman,
  Phys.\ Rev.\ E {\bf 59}, 364 (1999).
\bibitem{TeVo00} T. T\'el and J. Vollmer,
  in {\em Encyclopaedia of Mathematical Sciences}, Vol.\ 101,
  ed.: D. Sz\'asz (Springer Verlag, 2000), pp. 367-418.

\bibitem{KaT} G. K\'arolyi and T. T\'el,
  Physics Reports {\bf 290}, 125 (1997).

\bibitem{NeLoHeT} Z. Neufeld, C. L\'opez, E. Hern\'andez-Garc\'{\i}a,
  and T. T\'el,
  Phys.\ Rev.\ E {\bf 61}, 3857 (2000).

\bibitem{EckmRuel85} J.-P. Eckmann and D. Ruelle, 
                Rev.\ Mod.\ Phys.\ {\bf 57}, 617 (1985).
\bibitem{Gasp96} P. Gaspard,  Phys.\ Rev.\ E {\bf 53}, 4379 (1996).
\bibitem{Buni85} L. A. Bunimovich,
  Sov.\ Phys.\ JETP {\bf 62}, 842 (1985)
\bibitem{PiYo79} G. Pianigiani and J. A. Yorke,
  Trans.\ Am.\ Math.\ Soc.\ {\bf 252}, 351 (1979);
\bibitem{PiYo81} G. Pianigiani, J. Math.\ Anal.\ Appl.\ {\bf 82}, 75 (1981).
\bibitem{Tel90} T. T\'el, in
  {\em Directions in Chaos}, Vol.\ 3, edited by Hao Bai-Lin
  (World Scientific, Singapore, 1990), pp.\ 149-211.
\bibitem{Rugh92} H.H. Rugh, the correlation spectra for chaotic maps,
                 PHD. thesis, Niels Bohr Institute, unpublished (1992).
\bibitem{KaGr85} H. Kantz and P. Grassberger, Physica D {\bf 17},
         75 (1985).
\bibitem{Gasp92} P. Gaspard, J. Phys. A, Math. Gen. {\bf 25}, L483 (1992).
\bibitem{NeSz95} A. N\'emeth and P. Sz\'epfalusy,
  Phys.\ Rev.\ E {\bf 52}, 1544 (1995).
\bibitem{LuSz96} H. Lustfeld and P. Sz\'epfalusy,
  Phys.\ Rev.\ E {\bf 53}, 5882 (1996).
\bibitem{GrFu82} S. Grossmann and H. Fujisaka, Phys Rev {\bf A 26},  
         1179 (1982).
\bibitem{GeNi84} T. Geisel and J. Nierwetberg, Z. Phys. {\bf B56},
         59 (1984).
\bibitem{KLNSz} Z. Kaufmann, H. Lustfeld, A. N\'emeth, and P. Sz\'epfalusy,
  Phys.\ Rev.\ Lett.\ {\bf 78}, 4031 (1997).
\bibitem{CMar} P. Collet and S. Martinez,
  Nonlinearity {\bf 12}, 445 (1999).
\bibitem{ClGa} I. Claus and P. Gaspard,
  Phys.\ Rev.\ E, {\bf 63}, 036227 (2001)
\end{thebibliography}
\end{document}